\documentclass[conference]{IEEEtran}
\IEEEoverridecommandlockouts
\usepackage{amsmath,amssymb,amsfonts}
\usepackage{algorithmic}
\usepackage{graphicx}
\usepackage[hyphens]{url}  
\usepackage{xcolor}
\usepackage{float}
\usepackage{caption} 
\usepackage{booktabs} 
\usepackage[style=ieee,dashed=false]{biblatex}
\usepackage[hidelinks]{hyperref}
\usepackage[T2A,T1]{fontenc}
\usepackage[utf8]{inputenc}
\usepackage[russian,english]{babel}
\usepackage{csquotes}
\usepackage{tablefootnote}

\newcommand{\todo}[1]{}
\renewcommand{\todo}[1]{{\color{red} TODO: {#1}}}
\def\BibTeX{{\rm B\kern-.05em{\sc i\kern-.025em b}\kern-.08em
    T\kern-.1667em\lower.7ex\hbox{E}\kern-.125emX}}

\usepackage{tikz}
\newcommand*\circled[1]{\tikz[baseline=(char.base)]{
            \node[shape=circle,draw,inner sep=2pt] (char) {#1};}}
\begin{document}

\title{Showing the Receipts: Understanding the Modern Ransomware Ecosystem}

\author{
\IEEEauthorblockN{Jack Cable}
\IEEEauthorblockA{
\textit{Independent Researcher}\\
}
\and\IEEEauthorblockN{Ian W. Gray}
\IEEEauthorblockA{
\textit{New York University}\\
}
\and
\IEEEauthorblockN{Damon McCoy}
\IEEEauthorblockA{
\textit{New York University}\\
}
}


\maketitle

\begin{abstract}
Ransomware attacks continue to wreak havoc across the globe, with public reports of total ransomware payments topping billions of dollars annually. While the use of cryptocurrency presents an avenue to understand the tactics of ransomware actors, to date published research has been constrained by relatively limited public datasets of ransomware payments.

We present novel techniques to identify ransomware payments with low false positives, classifying nearly \$700 million in previously-unreported ransomware payments. We publish the largest public dataset of over \$900 million in ransomware payments -- several times larger than any existing public dataset. We then leverage this expanded dataset to present an analysis focused on understanding the activities of ransomware groups over time. This provides unique insights into ransomware behavior and a corpus for future study of ransomware cybercriminal activity.
\end{abstract}


\section{Introduction}

The rise of Ransomware as a Service (RaaS) in recent years has underpinned a significant increase in the scale and sophistication of ransomware operations. RaaS, a scheme in which ransomware operators lease ransomware kits to affiliates in exchange for a portion of proceeds, has facilitated a shift from a relatively low number of operators and opportunistic targeting of individual devices to robust operations with attacks tailored to their victims~\cite{oosthoek2022tale}. RaaS cybercriminals price ransom demands according to their victims' revenue and often engage in "big game hunting" to target entities that are likely to pay larger sums to recover their data. Researchers have documented this shift in organization and targeting, along with the increasing revenue of RaaS groups, through analysis of crowdsourced ransomware payments and public datasets~\cite{oosthoek2022tale}.

Measuring the total scale of the ransomware payments market is key to understanding -- and, ideally, disrupting -- the ransomware economy. Historically, public datasets of ransomware payments have lagged behind proprietary analysis tools, which often have direct access to reports of ransomware payments. Chainalysis, a blockchain analytics firm, started publicly reporting ransomware payments in 2019~\cite{ransomware_mass_market}. According to their most recent estimates, ransomware payments totaled \$220 million in cryptocurrency that year, \$905 million in 2020~\cite{RansomwareSkyrocketed2020}, \$983 million in 2021~\cite{Chainalysis2021Crypto}, \$567 million in 2022~\cite{RansomwareRevenueMorea}, and \$1.1 billion in 2023~\cite{RansomwareHitBillion}.

These estimates were assessed through blockchain analysis, proprietary software, and collaboration with law enforcement agencies, financial institutions, and incident response companies. However, industry reports on ransomware payments do not provide details on research methodology to identify ransomware payments on the blockchain, nor do they publish the underlying payment addresses. To help bridge this gap, Ransomwhere, the largest public dataset of ransomware payments, has cataloged \$278 million in ransomware payments at the time of publication~\cite{ransomwhere}. Larger and more representative open datasets of ransomware payments would help facilitate more comprehensive studies of the ransomware ecosystem.

To date, approaches for identifying ransomware payments have largely relied on either direct reports of payments or generalized blockchain clustering techniques. In this paper, we develop a novel framework for identifying and measuring ransomware payments. Leveraging a unique characteristic of the ransomware ecosystem, ransomware negotiators, we develop heuristics to classify previously unidentified ransomware payments. By tracing ransomware payments at their source -- where negotiator activity makes it possible to identify numerous payments from the same negotiator -- we can gain a deeper understanding of the size and scale of the ransomware ecosystem.

Through these heuristics, we expose over \$700 million in ransomware payments. Our validation indicates that our framework has a low false positive rate and can identify ransom payments at a larger scale than previous research. As such, our heuristics-based approach drives key insights into the activities of RaaS groups and their affiliates, including overlap and rebranding. We publish our dataset of \$900 million in ransomware payments\footnote{\url{https://github.com/cablej/showing-the-receipts}} -- nearly 4 times as much as previously published research -- to pave the way for future research that further assesses the ransomware ecosystem.

Our analysis confirms previously-documented trends in proprietary reports, such as that the average ransomware payment steadily increased between 2019 and 2022 (see Figure~\ref{fig:paymentsOverTime}). Additionally, we derive labels of ransomware payments and undertake an analysis of ransomware families. Based on common destinations of funds, we observe ties between multiple ransomware families consistent with prior proprietary reports (see Section~\ref{families}). Our proposed shared exposure methodology presents a way to understand links between ransomware operators and their affiliates through blockchain analysis. We also study splitting behavior, by which ransomware operators and affiliates divide their funds. We observe unique splitting rates per ransomware family and that the percentage kept by affiliates tends to increase as the amount of the ransom payment increases.

Taken together, we paint a picture of the ransomware ecosystem at a scale that is an order of magnitude larger than any prior published academic research. Our published dataset offers a strong starting point for additional research to study the ransomware ecosystem to the tune of nearly a billion dollars.

\section{Background}
\label{background}

Due to both a drive for profitability and greater scrutiny from law enforcement, ransomware actors have grown increasingly mature in their tactics, techniques, and procedures. These groups have adopted obfuscation techniques like multi-stage laundering and developed specialized roles within their cybercriminal organizations. Ransomware actors continually adapt their methods to evade detection and maintain profitability in the face of improving blockchain analysis and law enforcement action. In this section, we provide an overview of the stakeholders and their roles in this ecosystem. We then explain the steps of paying a ransom, including the role of negotiators in recovering from a ransomware incident. Finally, we describe the current state of ransom payment analysis. 

\subsection{Ransomware Stakeholders}
\begin{figure}[t]
\centering\includegraphics[width=8cm]{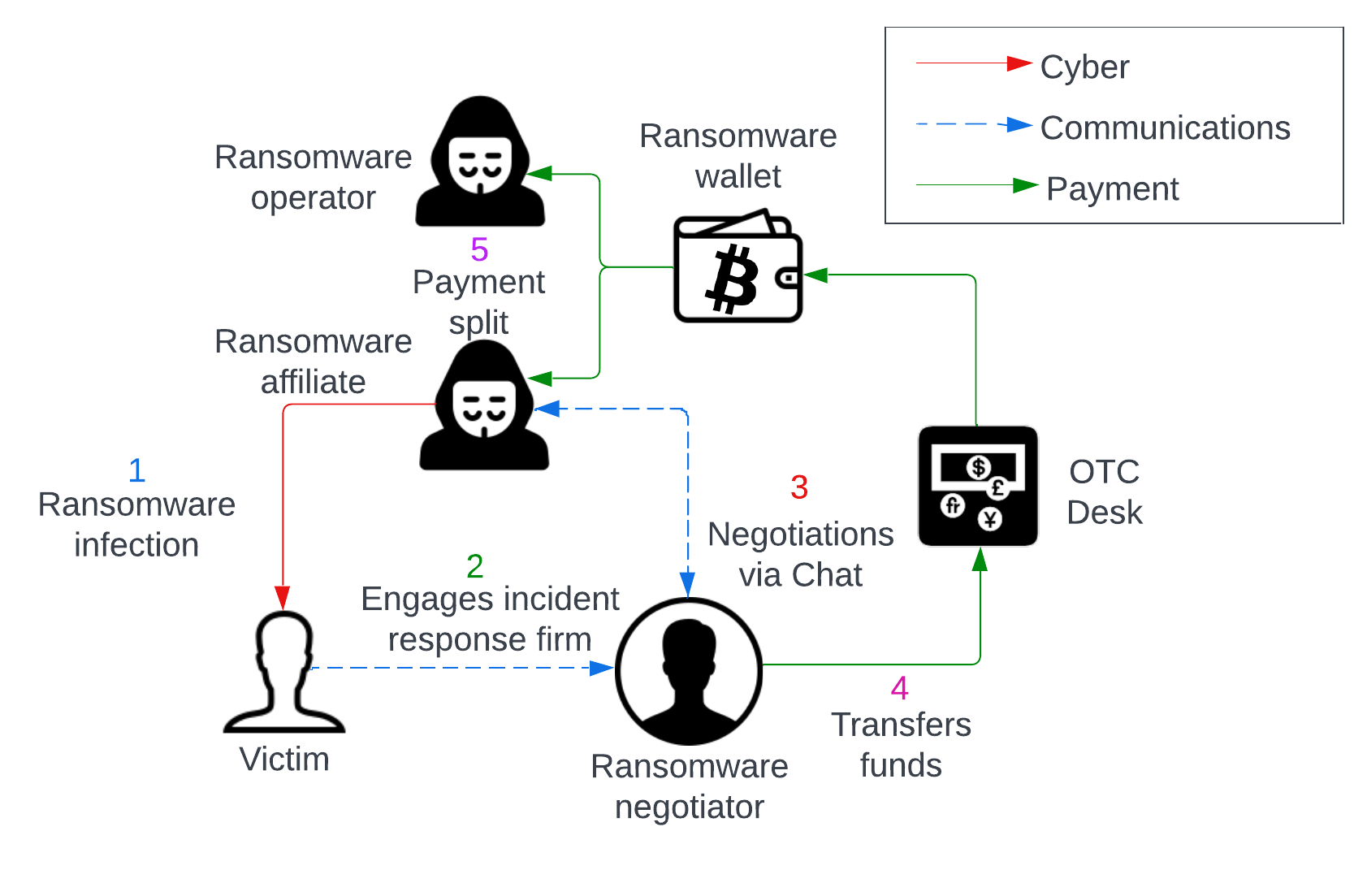}
\caption{Ransomware payment process.}
\label{fig:Ransomware_Diagram}

\end{figure}

As Figure~\ref{fig:Ransomware_Diagram} shows, there are often many stakeholders involved in the ransomware payment process. We briefly describe each of the core stakeholders, focusing on the activities of RaaS groups and their affiliates.

\noindent\textbf{Ransomware Operator} The ransomware operator specializes in development and operation of the ransomware. Operators often outsource the infection and payment negotiations to a third-party affiliate that leases the ransomware, and works on commission. Upon receiving the ransom payment, the operator splits the payment with the affiliate, with the operator typically receiving a smaller portion of the ransom payment. The percentage varies based upon RaaS group, the relationship between the operator and affiliate, and ransom amount. 

\noindent\textbf{Ransomware Affiliate} The affiliates specialize in gaining access to victim networks, deploying the ransomware, negotiating with the victim, receiving payments, and profit sharing. They split the ransom payment with the operator and normally receive a majority of the ransom. Some affiliates are able to arrange for a larger percentage of the ransom payment. Affiliates may freely move between ransomware groups, based on commission, ransomware type, or availability. 

\noindent\textbf{Victim} The victim is the organization that is infected with ransomware. Victims must determine the best method to recover their access and data, whether by restoring through backups or paying a ransom. Larger revenue organizations often have insurance or emergency response services contracts. Most victims work with a third-party incident response team to help guide them on how to handle their response. If the victim decides to explore paying the ransom they often hire a third-party company to negotiate with the ransomware affiliate.

\noindent\textbf{Ransomware Negotiator} A ransomware negotiator is a third party that navigates the high-stakes process of communicating with ransomware attackers after an infection. Their primary responsibilities include establishing communication with the attackers, gathering critical information about the ransom demands and attack specifics, assessing available options such as decryption tools or backups, and strategically negotiating the ransom amount and payment terms. If the victim decides to pay the ransom, the negotiator can purchase the cryptocurrency from a third-party exchange and execute the payment~\cite{RansomwareNegotiations2021}.

\noindent\textbf{OTC Desk} Larger ransomware negotiators likely operate or work with cryptocurrency over-the-counter (OTC) desks that facilitate large-scale purchases of cryptocurrencies directly between buyers and sellers, often outside traditional exchanges. These desks cater primarily to institutional investors, high-net-worth individuals, and entities seeking to trade significant amounts of cryptocurrencies without impacting market prices. OTC desks offer advantages such as reduced slippage, the difference between the expected price of a cryptocurrency trade and the actual price when the trade is executed, compared to trading on public exchanges~\cite{MediumOTC2020}. 

\subsection{Ransomware Payment Process}
Having established the core stakeholders that are involved in a ransomware attack, we now proceed to describe the typical ransomware payment process. Figure~\ref{fig:Ransomware_Diagram} shows a generalized flow of ransomware payments. 

A victim organization will typically become aware of a ransomware infection through a ransom note on the victim's screen, an extension on their files indicating that they have been encrypted, or slowdowns or performance issues. After an affiliate infects an organization with ransomware~\circled{1}, if the victim chooses to explore paying a ransom they will typically engage with a ransomware negotiator as part of their incident response~\circled{2}. The ransomware negotiator will communicate on behalf of the victim through the ransomware actor's preferred communications platform, often a payment portal on their site~\circled{3}. Ransomware negotiators have relationships with cryptocurrency exchanges or OTC desks to transfer the funds to pay the extortion~\circled{4}. Once the payment is received, the ransomware operator will split the payment between the affiliate and the operator~\circled{5}, with the affiliate normally receiving the larger percentage. Several industry and academic publications have detailed their observations of the ransomware payments ecosystem. 

\subsection{Ransomware Payments Analysis} \label{subsec:ransomware-payments-analysis}
Blockchain analysis firms such as Chainalysis publish estimation of ransomware campaign revenue in annual reports~\cite{RansomwareHitBillion}. These firms began publishing such reports in 2019, coinciding with the rise of RaaS, capturing the growth of financial crimes paid in cryptocurrencies. In 2019, blockchain analysis firms reported that ransomware attacks increased sharply, however the ecosystem at the time was predominantly commodity ransomware and smaller RaaS groups. The total ransomware payments in 2019 reached \$220 million ~\cite{RansomwareGoesMassa}. Multiple cybercrime groups shifted to the RaaS model following the introduction of ransomware leak sites and double extortion ~\cite{RansomwareSkyrocketed2020}. In 2020, ransomware was reported to increase to \$350 million, though updated figures later showed payments reaching \$983 million, likely as a result of victim underreporting ~\cite{RansomwareSkyrocketed2020}. In 2021, reported ransomware payments exceeded \$983 million, observing news strains and an increase in active strains~\cite{Chainalysis2021Crypto}. In 2022, ransomware payments declined to \$567 million, likely influenced by Russia's invasion of Ukraine, and law enforcement interventions~\cite{RansomwareRevenueMorea}. In 2023, reported ransomware payments exceeded \$1 billion, observing record-breaking payments and an increase in the scope and complexity of ransomware attacks~\cite{RansomwareHitBillionb}. Blockchain analysis firms do not publish cryptocurrency addresses associated with ransomware or other cryptocrimes, as they may continue to work with law enforcement on investigations~\cite{chainalysisOFACSanctions}.

From an academic standpoint, the lack of transparency in ransomware payments presents significant challenges in understanding the scope and impact of these cybercrimes. As previously highlighted, ransomware payments are likely underreported as victims may not wish to disclose a security incident~\cite{cybersecuritydiveHiveTakedown}. Ransomware payment addresses are often not openly shared, which creates a barrier to additional analysis from a broader community into key questions such as understanding the victims, payment amounts, or intermediaries facilitating ransomware transactions. To close this gap, the Ransomwhere dataset~\cite{ransomwhere}, established in 2021, accepts anonymous, crowdsourced reports of ransomware payments. Several research studies have used this dataset to understand ransomware payments, assess laundering patterns, characterize ransomware markets, and develop heuristics to better identify additional addresses and trace transactions for specific ransomware families~\cite{oosthoek2022tale, Gray2022MoneyOM}. Even then, the scale of public datasets like Ransomwhere is significantly lower than the numbers reported by proprietary analysis firms.

Prior studies that have added to the public set of likely ransomware payments have highlighted the occurrence of payment splitting between ransomware operators and their affiliates. For example, DarkSide's revenue split reportedly varied upon the revenue of the victim company~\cite{congAnatomyCryptoEnabledCybercrimes2022a}. Research into the Conti Ransomware identified that split percentages varied from 5\% to 40\%, with most addresses splitting at a fixed 20\% rate~\cite{Gray2022MoneyOM}. Gautschi analyzed LockBit payments on the blockchain and observed a split involving one input and two outputs, with administrators receiving 20\% and affiliates retaining 80\%, which is consistent with LockBit's advertised splitting rate~\cite{gautschiFindingAnalyzingLockbit}. We build on these studies of specific groups by providing an analysis of splitting behavior across many ransomware families based on our payment dataset. 

Several research papers have assessed the scale of ransomware payments, either through the analysis of single ransomware families or across the ecosystem at large. Paquet-Clouston et al. researched 15 separate strains of commodity ransomware from 2013-2017, identifying \$12,768,536 of ransomware payments~\cite{paquet-cloustonRansomwarePaymentsBitcoin2019a}. Conti et al. conducted a study on the economic impact of ransomware campaigns and assessed CryptoLocker payments at \$42,292,191.17, and CryptoWall at \$45,370,589 ~\cite{contiEconomicSignificanceRansomware2018a}. Oosthoek et al. studied the Ransomwhere dataset, providing analysis of ransomware payments across 87 ransomware families composing \$101 million~\cite{oosthoek2022tale}. Our expanded dataset of ransomware payments allows us to study payments at a scale an order of magnitude larger than any existing work, providing a more holistic analysis of this ecosystem.
\section{Data}
\label{data}

\subsection{Input Datasets}

We utilize three primary data sources to inform our research methodology: the public Bitcoin blockchain, the open-source Ransomwhere dataset,\footnote{\url{https://ransomwhe.re/}} and labels from Crystal,\footnote{\url{https://crystalintelligence.com/}} which is a proprietary blockchain analysis product. Additionally, we leverage data from another proprietary blockchain analysis firm as part of our validation approach.

We operate a Bitcoin full node, which we use to collect all Bitcoin transactions up to February 2024 at a block height of 830,956. As described below, we use this public blockchain data to expand and analyze two clusters that appear to be responsible for originating large amounts of ransomware payments. We utilize the open source blockchain analysis tool BlockSci with its default multi-input clustering techniques to extract and analyze Bitcoin blockchain data~\cite{blocksci}.

Next, to obtain known "in the wild" ransomware payment addresses, we leverage the public Ransomwhere dataset~\cite{ransomwhere}. At the time of publication, Ransomwhere is a public dataset of 10,454 ransomware payment addresses representing \$278 million in payments. Ransomwhere aggregates validated crowdsourced reports in addition to data from all previous major papers in the field with publicly-available addresses, as noted by Gomez et al.~\cite{gomezCybercrimeBitcoinRevenue2023b}. Ransomwhere's dataset includes payments from 2015 to 2023, although most payments in the dataset occurred prior to 2023. Given our paper's focus on modern ransomware activity dominated by RaaS groups, as opposed to older, commodity ransomware actors, we focus only on RaaS payments in the Ransomwhere dataset. We attribute 292 addresses representing \$240 million in payments to RaaS groups.

Lastly, we use labels from two proprietary blockchain analysis products, Crystal and an additional blockchain analysis firm. We utilize Crystal's labels (which include all addresses in the Ransomwhere dataset) as part of our heuristics to determine whether an address has sent payments to a known address associated with a ransomware group. We do not use the data from the additional blockchain analysis firm to develop our heuristics and instead use it as an independent source to validate our findings, as described in Section~\ref{validation}. We utilize both data sources as part of our analysis of ransomware families detailed in Section~\ref{analysis}.

\subsection{Output datasets}

We construct two datasets of ransomware payments, which we briefly describe here. The methodology by which we derived these datasets is described in Section~\ref{methodology}. As is standard in the literature~\cite{oosthoek2022tale}, we compute the approximate U.S. dollar (USD) value of a transaction by multiplying the Bitcoin (BTC) value of the transaction by the closing BTC-USD exchange rate the day the transaction occurred.

The first dataset is composed of 465 likely ransomware payment addresses which have collectively received \$401 million in payments. These payments span from August 2019 to December 2022. We identified these payments by tracing transactions from a single origin cluster. In practice, it is likely that this identified cluster was operated by a ransomware negotiator who processes payments on a victim's behalf. We note that these methodologies also identified 76 payments worth \$115 million that were already present in the Ransomwhere dataset. We do not include these payments in the new dataset, though this further validates the efficacy of our methodology in finding legitimate ransomware payments.

The second dataset has 256 likely ransomware payment addresses that have collectively received \$276 million in payments, spanning from June 2016 to December 2023. These payment addresses were obtained by our approach of traversing the graph of seed ransomware payments (from Ransomwhere and the first dataset) to identify neighboring payments that exhibited characteristics unique to ransomware payments, as described in Section~\ref{methodology}.

We publish both datasets on GitHub.\footnote{\url{https://github.com/cablej/showing-the-receipts}} Our dataset is, at time of publication, the largest public dataset of ransomware payments and, when combined with the existing Ransomwhere dataset (which has \$240M in RaaS payments at the time of publication), constitutes approximately \$917 million in likely ransomware payments.

Table~\ref{tab:data_overview} presents an overview of the datasets that we analyze in the remainder of this paper.

\begin{table}[t]

\caption{Overview of the datasets studied in this paper.}
\label{tab:dataset_overview}
\scalebox{0.87}{

\begin{tabular}{lccc}
\toprule
Dataset          & Num. Addresses & Total Received & Date Range              \\ \midrule
Ransomwhere (RaaS)      & 292             & \$240M         & 01/2016 - 11/2023 \\ 
Orig Cluster     & 465             & \$401M         & 08/2019 - 12/2022 \\
Expanded Set & 256             & \$276M         & 06/2016 - 12/2023 \\ \midrule
Total  & 1,013            & \$917M       & 01/2016 - 12/2023 \\ 
\bottomrule
\end{tabular}
\label{tab:data_overview}
}
\end{table}
\section{Methodology}
\label{methodology}

The core of our methodology lies in an understudied aspect of the ransomware ecosystem: ransomware negotiators. As described earlier, most victims engage with ransomware negotiators to interface with the ransomware actor and process the ransomware payment on the victim's behalf. Whereas ransomware actors are incentivized to cover their tracks after the payment, negotiators are less likely to take steps to obscure the steps that occur immediately before a payment. This presents an avenue to identify ransomware payments.

In practice, at least one major negotiator operated in a manner that allows their payments to be readily identified. In prior work, Gray et al. identified a Bitcoin cluster as being responsible for over 70\% of studied ransomware payments to Conti~\cite{Gray2022MoneyOM}, one of the largest ransomware-as-a-service groups. This research analyzed 666 cryptocurrency addresses identified within the Conti leaks, internal chat logs between the various members of the ransomware group leaked in February 2022. 

The researchers analyzed ransomware addresses in the Conti leaks dataset, as well as the addresses within the Ransomwhere dataset. Through their analysis, the researchers  identified an unlabeled cluster of addresses being responsible for a large portion of the Conti ransom payments. They then discovered additional ransomware payment addresses by looking for addresses that sent money to an address within the leaked dataset, exhibited splitting behavior according to an exact percentage, and received more than 99\% of its funds from a low risk exchange.

We expand on Gray et al.'s research to understand the scope of ransomware payments to all ransomware families originating from this cluster. Our analysis of this cluster indicates that it is likely two closely related clusters. We denote these clusters as Cluster A, characterized by the address \texttt{19JyAkHKh36sFduqK4hMsMZhU6ZDoLotW}, and Cluster B, characterized by the address \texttt{3DtLWACQNiVFaXQyMS57PjVir19FRY32Hf}.\footnote{For reproducibility, when referring to "clusters" we refer to the cluster generated via BlockSci under standard configuration~\cite{blocksci}.}  Cluster B has received the vast majority of its funds directly from Cluster A, suggesting that ownership is shared between these clusters. As we shall see in the subsequent analysis, it appears that the majority of transactions from Cluster B is a ransomware payment, while only a subset of the transactions from Cluster A are ransomware payments. We assess that ransomware payments originating from both clusters are likely related to the activity of a ransomware negotiator, and that Cluster A is likely housed within an OTC desk used by the negotiator.

Leveraging the Ransomwhere dataset, we first show that these clusters are responsible for originating a plurality of ransomware payments observed in the wild. This suggests that the presumed negotiator operating these clusters has a significant market share of overall ransomware payments.

As shown in Table~\ref{tab:sum_clusters}, cluster A, at the time of writing, has received 1,962,574 BTC across 32,840 transactions, although a large portion of these transactions are large payments to/from known exchanges (likely to fund the OTC desk). Cluster A was active from August 20, 2020 to October 7, 2023. Cluster B has received and sent 14,309 BTC over 363 transactions and was active from August 25, 2020 to March 5, 2023.

\begin{table}[t]
\caption{Summary of clusters A and B.}

\begin{center}
\centering
\small %
\begin{tabular}{cccc} 
 \toprule
 Cluster & BTC & Transactions & Range \\  \midrule
 
 A & {1,962,574} & {32,836} & 8/20/20 - 10/7/23 \\ 
 
 B & {14,309} & 363 & 8/25/20 - 3/5/23 \\
 \bottomrule
\end{tabular}
\end{center}
\label{tab:sum_clusters}
\end{table}

Notably, it would appear that payments from these clusters stopped entirely soon after the publication of Gray et al.'s paper. The last transaction from Cluster A was sent on September 22, 2023, while the last transaction for Cluster B was sent on March 5, 2023. If the clusters do indeed belong to a ransomware negotiator, this suggests that the negotiator might have either switched to a new cluster or developed more sophisticated techniques following the publication of their payments.\footnote{Some blockchain analysis tools suggest that these clusters were operated on Genesis Global Trading's OTC desk, which had a large amount of exposure in FTX and shuttered trading in September 2023~\cite{genesis}.}

\begin{figure}[t]
\centering\includegraphics[width=0.45\textwidth]{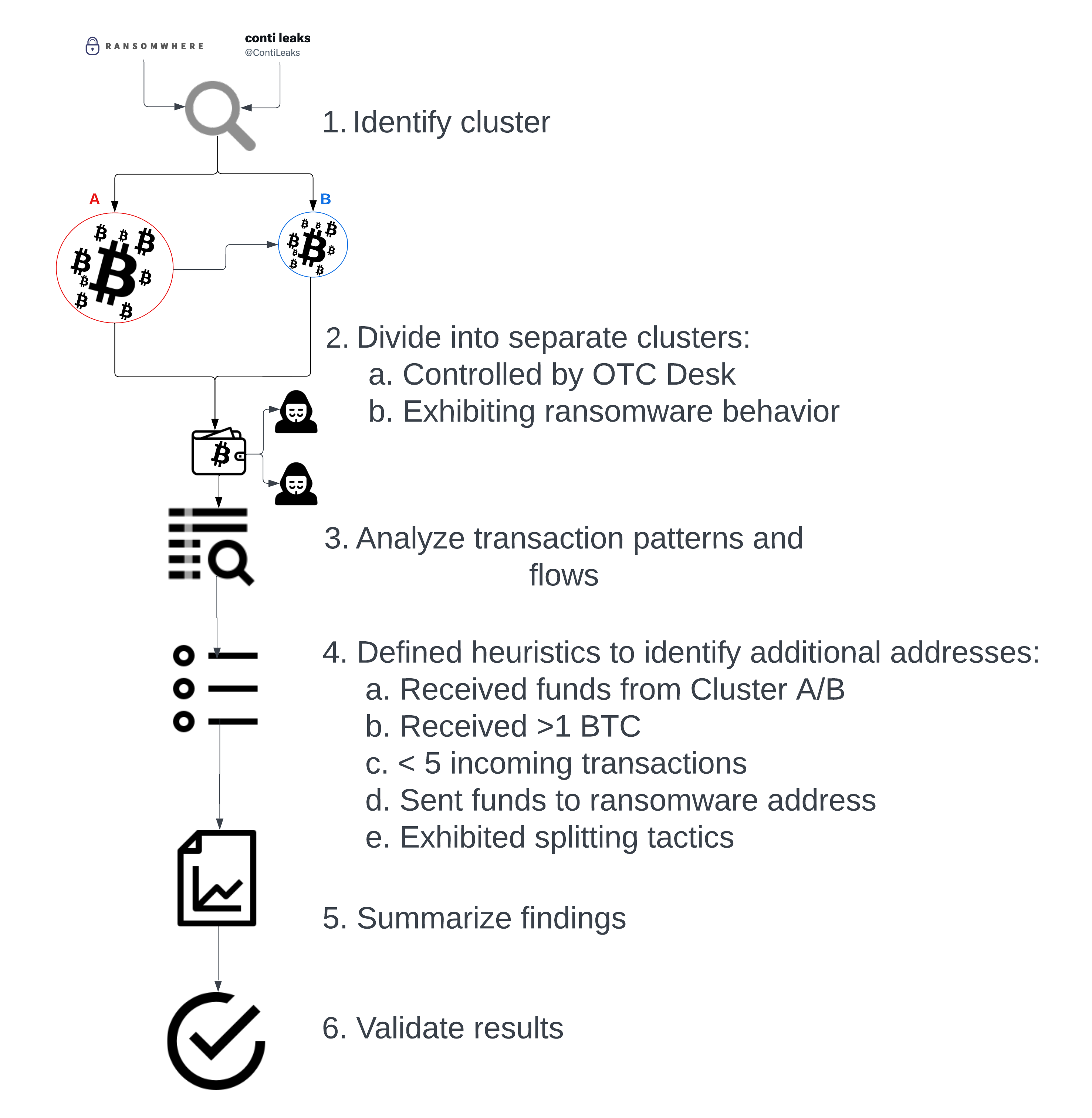}
\caption{Methodology for identification of ransomware payments.}
\label{fig:Methodology_Infographic}
\end{figure}

Figure~\ref{fig:Methodology_Infographic} provides and overview of the steps in our ransomware payment methodology which we describe in detail below.

\subsection{Cluster identification}

In order to identify the clusters that most often originate ransomware payments observed in the wild, we leverage known ransomware payments from the Ransomwhere dataset. We perform clustering using BlockSci and backtrace to a depth of up to 3 transactions before the ransomware payment.

To distinguish between modern RaaS and commodity ransomware, we filter the Ransomwhere dataset to addresses associated with RaaS groups. Of 292 RaaS addresses, the plurality of payments originated from Clusters A and B, which together represented 41\% of payments -- 37\% from Cluster A and 4\% from Cluster B. The next 3 top sources were the large cryptocurrency exchanges Gemini (22\%), Binance (18\%), and Coinbase (15\%).

This demonstrates that, of the largest public set of ransomware payments, a significant portion of them originate from Clusters A and B. Given the relative smallness of these clusters (the major exchanges, according to Crystal, have on the order of tens of millions of transactions rather than tens of thousands), ransomware payments represent a nontrivial portion of overall payments from these clusters, and we hypothesize that we can successfully identify additional ransomware payments by studying these clusters. An industry analysis of ransomware payments cites \$3.8 billion in payments since 2019~\cite{RansomwareHitBillion}, and assuming that the Ransomwhere sample is representative, it is possible that Clusters A and B have processed over a billion dollars in payments to ransomware actors.

We note that while Ransomwhere may not have a fully representative sample of ransomware transactions, it is still very likely true that these clusters represent a significant portion of ransomware payments made. In particular, bias in the Ransomwhere dataset can be introduced due to the nature of voluntary reporting, thus potentially over-representing the portion of transactions from Clusters A and B. The Ransomwhere dataset includes data from Gray et al.~\cite{Gray2022MoneyOM}, which identify Cluster A as being responsible for 70\% of payments to Conti. We observe that even removing Gray et al.'s attributed payments to Conti, a large portion -- 28\% -- of payments originate from these clusters.

\subsection{Heuristics and Payment Identification}

Having identified the two clusters responsible for a significant portion of ransomware payments, we now describe our methodology to identify ransomware payments from these clusters. We develop the following criteria empirically based on existing data:

\begin{enumerate}
\item The address directly received its funds from Cluster A or Cluster B;
\item The address received at least 1 BTC;
\item The address has at most 5 incoming transactions; and
\item At least one of the following is true:
\begin{enumerate}
\item The address sent funds to a known ransomware address; or
\item The address sent funds to an address that received funds from a known ransomware address and the address exhibited splitting tactics commonly associated with ransomware.\footnote{For these items, we leverage Crystal's database of ransomware address labels.}
\end{enumerate}
\end{enumerate}

These criteria encapsulate addresses that both stem from a source known to originate ransomware payments (on the incoming side), and then have a demonstrated connection to a known ransomware payment (on the outgoing side). The last heuristic captures a common tactic among ransomware actors, where they often pool payments together after receiving them~\cite{RansomwareRevenueMore}.

Criteria 2 and 3 ensure that the activities of the addresses are consistent with modern RaaS behavior, which typically leverage one address per victim and receives payments in the hundreds of thousands of dollars~\cite{oosthoek2022tale}. For instance, the Q1 2024 report from Coveware, an incident remediation company, observed the median ransom payment at \$250,000 (approximately 3.6 BTC)~\cite{coveware}.

Applying these criteria to all addresses that received payments from Clusters A and B yields 541 ransomware payments which have received a combined \$516 million in payments. Of these, 76 were already present in the Ransomwhere dataset,\footnote{Of these 76 payments we identified in the Ransomwhere payment, 41 were sourced from Gray et al.'s study.} resulting in 465 newly identified ransomware payment addresses, receiving a combined \$401 million in payments. We  describe our approach to validation in Section~\ref{validation}. We then analyze these addresses in more depth in Section~\ref{analysis}.

We note that these two heuristics represent a conservative approach to identifying ransomware payment addresses. In particular, we only account for when an address interacts with an address tied to a known ransomware payment in our seed dataset. It is very likely that the true amount of ransomware payments originating from this cluster is higher; however, to ensure accuracy we limit to the listed heuristics.

\subsection{Expanded Dataset Identification}

Up to this point, we focused only on addresses that have originated from Clusters A and B. While we have demonstrated that a significant amount of ransomware payments can be found there, as evidenced by the earlier analysis of Ransomwhere payments we know that approximately two-thirds of payments originate from other clusters.

We postulate that Clusters A and B are relatively unique in that it is possible to enumerate every transaction stemming from those clusters and apply heuristics to understand whether they are ransomware payments. For most exchanges, such an approach is not possible due to the sheer volume of transactions. In the course of our research, with the exception of Clusters A and B, we did not find any other notable clusters originating ransomware payments that were not exchanges.

Here, we present our methodology to further extend our dataset of discovered ransomware payments. In essence, we traverse the graph of payments that we have already identified to find previously unidentified payments that exhibit behavior consistent with ransomware payments. 

Our methodology is a generalization of Gray et al.'s heuristics developed to identify Conti payment addresses. Gray et al. consider an address to be a Conti ransom payment if it (1) sent money to an address affiliated with Conti, (2) exhibited splitting, and (3) received more than 99\% of funds from a low-risk exchange or Clusters A or B.

Our heuristics are as follows:

\begin{enumerate}
    \item The address sent a nontrivial portion of its funds to an address that received a nontrivial portion of its funds up to 3 hops away from an address in Ransomwhere or identified in the previous section; \footnote{We use the Ransomwhere RaaS addresses and the addresses identified in the previous subsection as our list of known ransomware addresses.}\textsuperscript{,}\footnote{We base our computation of funds overlap using the multi-hop Jaccard similarity metric defined in Section~\ref{analysis}.}
    \item The address exhibited splitting, using Gray et al.'s splitting algorithm; and
    \item The address received more than 99\% of its funds from a low-risk exchange (based on Crystal's labels) or Clusters A or B;
\end{enumerate}

To discover addresses that meet these heuristics, we trace all addresses that have received funds within 3 hops from the addresses we identified in the previous subsection and addresses labeled as a RaaS payment in the Ransomwhere dataset. This yields 256 payment addresses which have received a combined \$276 million in payments. We analyze these payments in Section~\ref{analysis}.

Based on labels from Crystal, 256 payments, 23\% originate from Gemini, 14\% from Clusters A or B, 7\% from Coinbase, and 5\% from Binance. The remaining 50\% of payment origins are either smaller exchanges or are unlabeled. These three exchanges are also the most common sources of payments for Ransomwhere addresses. We note that the lower proportion of payments originating from Clusters A or B is in part due to excluding payments that had already been identified as part of the original cluster. This is nonzero, however, as our expanded dataset builds off of the original cluster, so an address that has ties to a ransomware payment in the original cluster but not a labeled address in Crystal could be identified as a ransomware payment through our expanded heuristics.

\section{Validation}
\label{validation}

To validate our dataset, we leverage an additional third-party blockchain analysis firm, herinafter referred to as Analysis Firm B. The data from Analysis Firm B was not used to inform our methodology or datasets, and hence provides an independent viewpoint from which to validate our data.

We note inherent limitations with validating any such dataset of ransomware payments. As there is no universal ground truth dataset of ransomware payments, it is impossible to say with certainty whether all identified payments are actual ransomware payments.
Nonetheless, we have developed some heuristics and metrics to assess the accuracy of our identified payments by using the data from Analysis Firm B.

We search each address present in our datasets within the data from Analysis Firm B. For each address, we evaluate whether:

\begin{enumerate}
    \item The address or portions of its payments are sent to known ransomware groups, suggesting the address is likely a ransomware payment (suspected true positive);
    \item The address or portions of its payments are sent to known illicit entities, suggesting the address is likely a ransomware payment (suspected true positive);
    \item The majority of payments are sent to known low-risk entities, suggesting the address is likely not a ransomware payment (suspected false positive).
    \item The destinations of all payments are unlabeled in Analysis Firm B, providing no information. 
\end{enumerate}



Of the 1,013 addresses, 521 (51\%) have sent all their funds to addresses labeled as ransomware, while 769 (76\%) have sent some portion of their funds to addresses labeled as ransomware within Analysis Firm B. Additionally, 577 (57\%) have sent all their funds to addresses labeled by Analysis Firm B as high-risk, while 954 (94\%) have sent some portion of their funds to addresses labeled as high-risk. High-risk addresses are those associated with illicit services, such as exchanges that have been known to enable criminal activity.

\begin{figure}[t]

\centering\includegraphics[width=8cm]{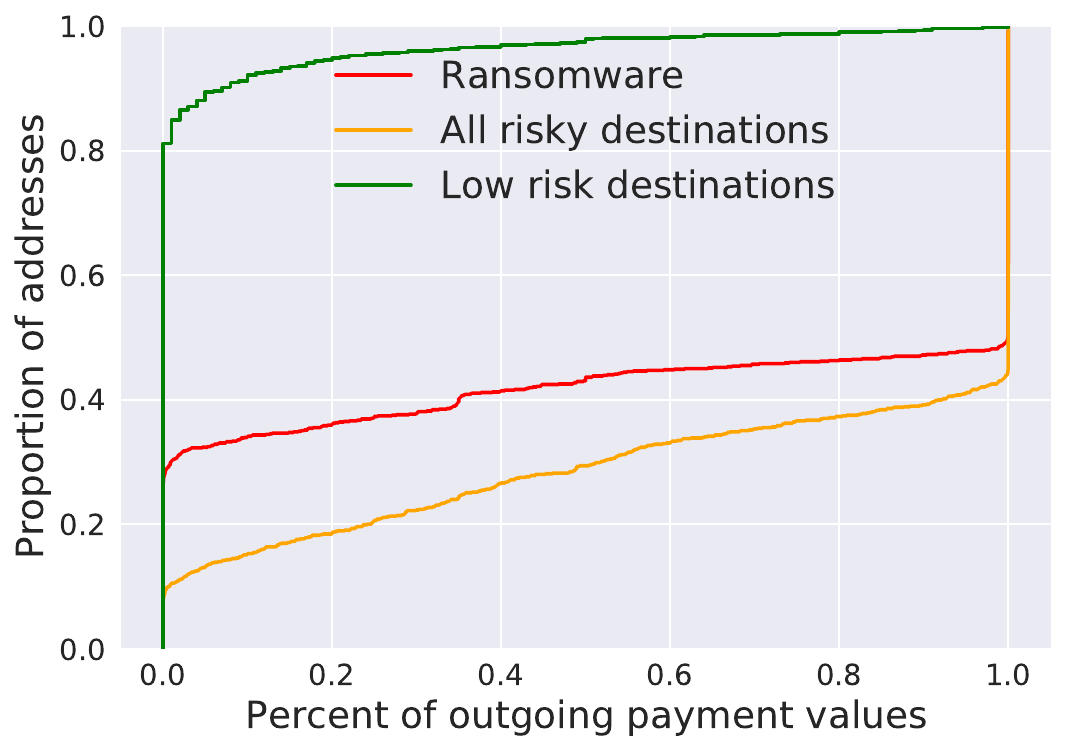}
\caption{ECDF of the percent of outgoing payment values labeled by Analysis Firm B as ransomware, high-risk, and low-risk.}
\label{fig:ecdf_trm}

\end{figure}

Of the remaining 59 addresses, 8 have sent a majority of funds to low-risk entities while 51 have a minority of destinations labeled in Analysis Firm B as low-risk and the remaining unlabeled. The ECDF of payment destinations within their data is shown in Figure~\ref{fig:ecdf_trm}. Notably, the range of payments that are partially labeled as ransomware or a risky destination -- approximately a quarter of payments for ransomware and a third of payments for risky definitions -- indicate that a large percent of payments may be newly discovered.

Table~\ref{tab:dataset_accuracy} shows our assessment of the accuracy of the three studied datasets. We observe that, unsurprisingly, nearly all payments in the Ransomwhere dataset are labeled as ransomware or otherwise illicit in the Analysis Firm B data. While both the original cluster and the expanded set have a high percentage of illicit destinations, addresses in the original cluster have a significantly higher portion of destinations labeled as ransomware. Over three-quarters of the original cluster's addresses have some ties to known ransomware addresses, and the vast majority have some ties to illicit activity. While a smaller portion -- approximately one-half -- of the expanded set have ties to known ransomware addresses, we see that nearly 90\% of these addresses still send funds to illicit destinations.

\begin{table}[t]
\caption{Destinations of payments by dataset. We consider an address to be "all ransomware" or "all illicit" if it has sent greater than 99\% of its funds to an address labeled by Analysis Firm B as ransomware or another illicit category.}

\label{tab:dataset_accuracy}
\scalebox{0.87}{
\begin{tabular}{lcccc}
\toprule
Dataset          & All Ransomware & Some Ransomware & All Illicit & Some Illicit \\ \midrule
Ransomwhere      & 89\%            & 91\%           & 99\%         & 99\%        \\
Orig Cluster     & 46\%            & 76\%           & 50\%         & 95\%        \\
Expanded Set & 19\%            & 58\%           & 23\%         & 87\%        \\ \midrule
Total   & 51\%            & 76\%           & 57\%         & 94\%        \\ \bottomrule
\end{tabular}
}
\end{table}

These results suggest that our methodology identifies ransomware payments with high precision. In particular, we note the low suspected false positive rate (8 of 1,013 addresses). These 8 addresses, 2 of which are from the original cluster and 6 are from the expanded set, total \$3.6 million, or 0.4\% of our dataset's total payments. Even then, we note that an address sending a majority of its funds to low-risk entities does not definitively rule out the possibility that the address is a ransomware payment.  Additionally, 51\% of addresses we identified are labeled in the Analysis Firm B data as ransomware, while only a fraction of these were publicly known previously, suggesting that our methodology can identify ransomware payments that previously were only known by proprietary tools. Furthermore, the additional 43\% of addresses that have some ties to illicit activities but are not fully labeled as ransomware payments by Analysis Firm B suggests that our methodology can identify ransomware payments that even cutting-edge tooling has not identified to date.





\section{Analysis}
\label{analysis}

We now proceed to analyze our constructed dataset of ransomware payments. We first present overall characteristics of the ransomware payments economy, and then explore three different aspects of ransomware payments activities: splitting, in which ransomware actors divide payments between core operators and affiliates, destination pools of ransomware payments, and overlap between ransomware groups. Together, these characteristics paint a picture of unique tactics, techniques, and procedures (TTPs) of ransomware families.

\subsection{Ransomware Ecosystem Analysis}

\begin{figure}[H]
\centering\includegraphics[width=8cm]{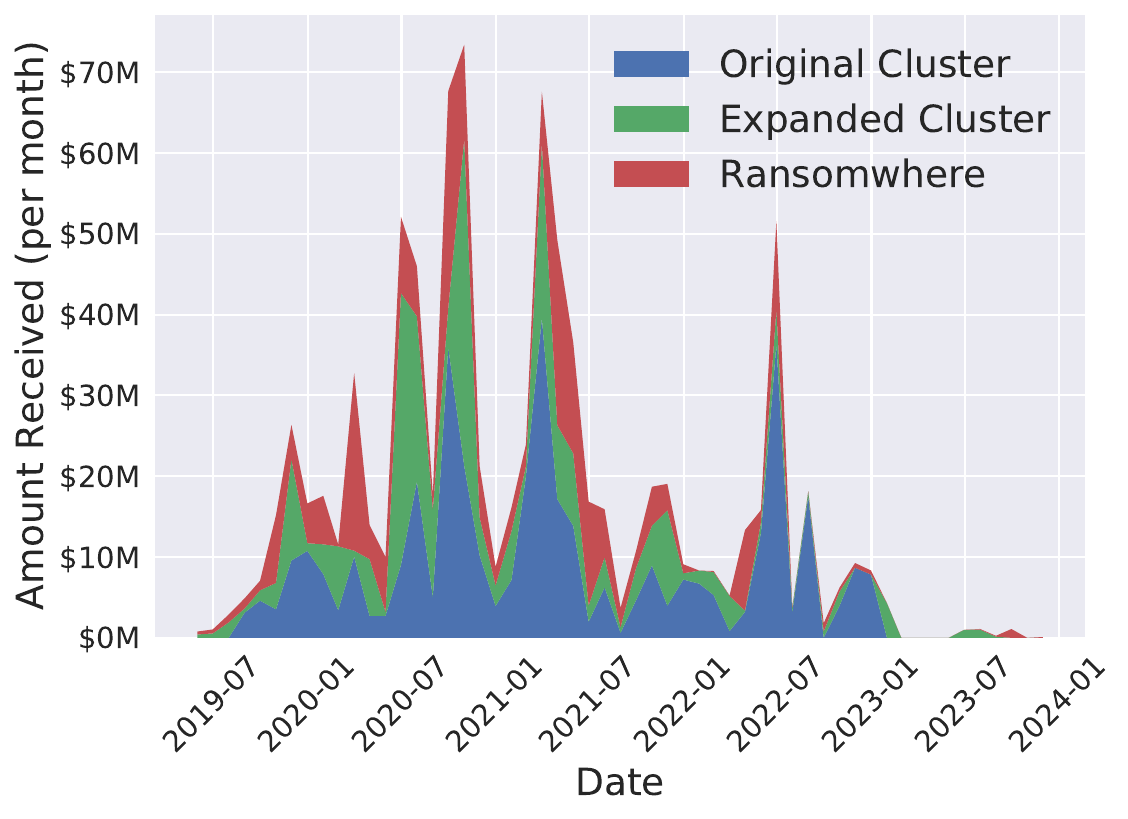}
\caption{Ransom payments received over time}
\label{fig:paymentsOverTime}
\end{figure}

To start, we review some overall metrics of our dataset. As described earlier in Table~\ref{tab:dataset_overview}, we have observed \$917M in payments across 1,013 addresses. The mean payment amount is \$905K and the median payment amount is \$254K. Figure~\ref{fig:paymentsOverTime} shows the amount of payments over time. Spikes in ransomware payments are visible in late 2020, early 2021, and mid 2022, which coincide with, for instance, reported spikes in ransomware payments amidst the start of the Covid pandemic~\cite{10308425}. We note that due to Clusters A and B ceasing operations in 2023, the majority of payments in our dataset occur prior to 2023.

\begin{figure}[H]
\centering\includegraphics[width=0.45\textwidth]{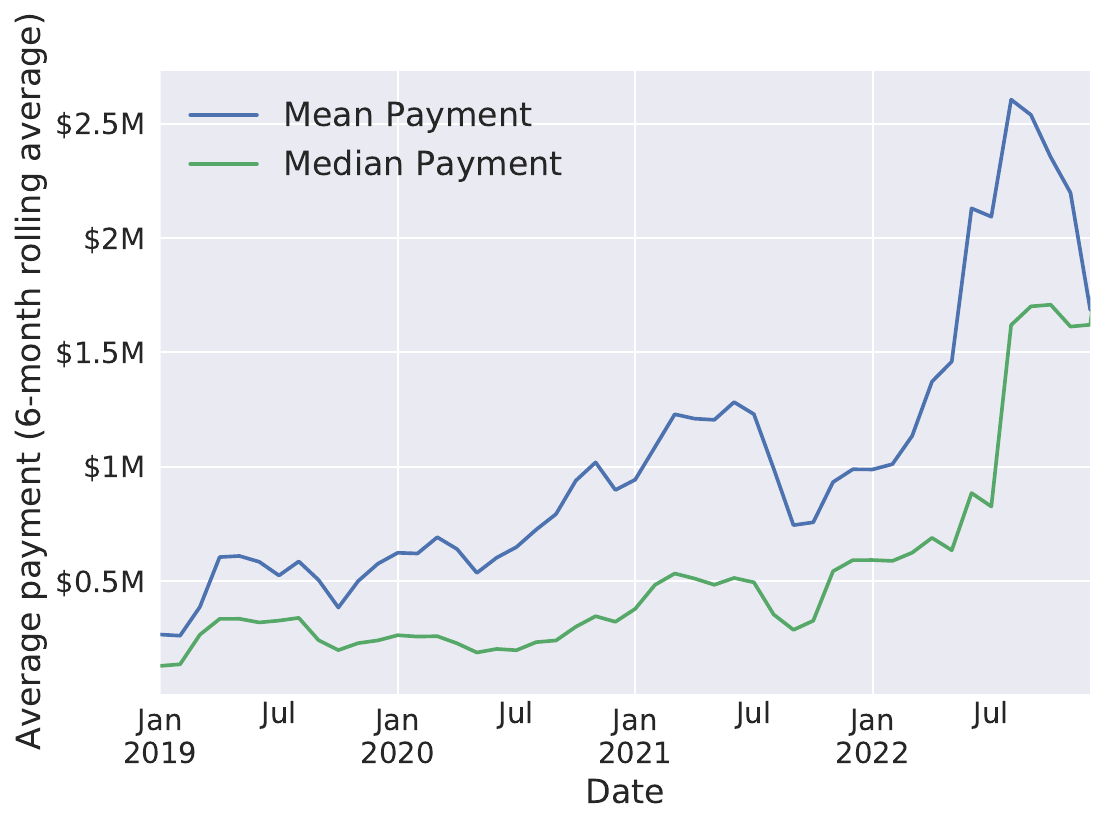}
\caption{Mean and median ransom payments over time.}
\label{fig:meanPayments}
\end{figure}

Our dataset also confirms previously-reported trends, including the trend of ransomware payments increasing over time. Figure~\ref{fig:meanPayments} shows the mean and median ransom payments over time. We observe a steady increase in the mean ransomware payment from approximately \$250K in 2019 to \$1M in 2021 and spiking to \$2.5M in 2022. This mirrors trends in other public reports, such as Coveware's~\cite{coveware}. We note that, despite similar trends, Coveware's reported average ransomware payment is significantly lower, suggesting that Coveware may have additional insight into lower-amount ransom payments. Our dataset also exhibits similar trends such as the increasing percentage of ransomware payments over \$1 million over time reported by Chainalysis~\cite{RansomwareHitBillion}. We observe an increase in the percent of million-dollar payments from approximately 10-20\% in 2019 to 50-60\% in 2022.

\begin{figure}[t]
\centering\includegraphics[width=9cm]{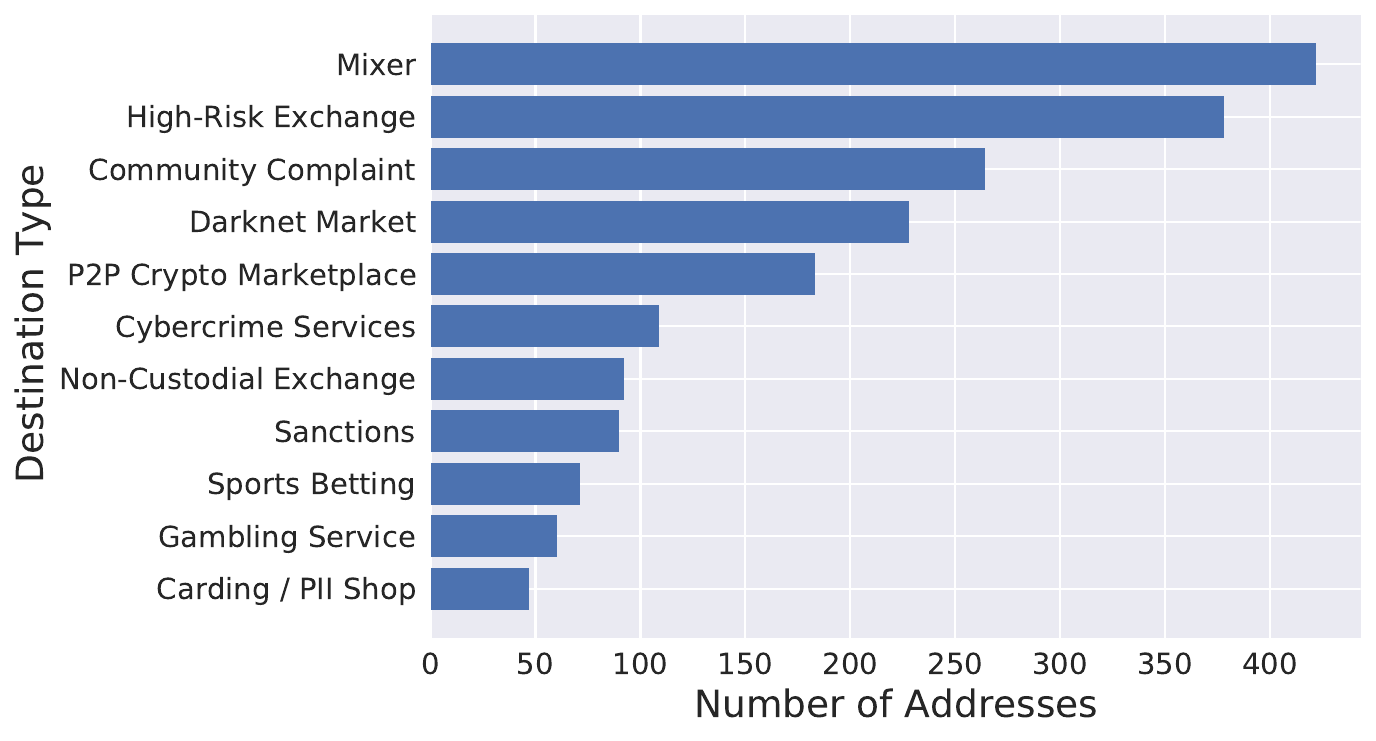}
\caption{Top 10 most common illicit destination types, from the blockchain analysis firm data.}
\label{fig:bafIllicitDests}
\end{figure}

Leveraging labels from the additional blockchain analysis firm, we identify the most common illicit destination of ransomware payments, depicted in Figure~\ref{fig:bafIllicitDests}. Notably, while mixers are the most common destination, only 42\% of addresses send to mixers. This is consistent with previous research~\cite{oosthoek2022tale} and suggests that even RaaS groups do not consistently launder payments through mixers. This enables in-depth analysis, such as studying the splitting behavior of RaaS groups.

\subsection{Family Analysis}
\label{families}

To enrich our dataset, we assign labels of ransomware families to addresses in our original cluster and expanded datasets leveraging the data from the blockchain analysis firm and Crystal. We label a ransomware payment as the family it has sent the most funds to, as labeled by the blockchain analysis firm or Crystal. If the address has not sent funds to any address labeled as ransomware by the blockchain analysis firm or Crystal, we leave the address unlabeled. For Ransomwhere addresses, we use the existing labels in the dataset. Table~\ref{tab:mostCommonFamilies} shows the most common ransomware families.



\begin{table}[t]
\caption{The 10 most common ransomware families in our dataset.}
\label{tab:mostCommonFamilies}
\scalebox{0.9}{
\begin{tabular}{lrrrr}
  \toprule

Family       & \# Ransomwhere & \# Orig Cluster & \# Expanded & \# Total \\
  \midrule

Conti        & 102            & 130 & 80   & 312      \\
NetWalker    & 66             & 84  & 17   & 167      \\
Ryuk       & 25             & 9     & 7 & 41       \\
MedusaLocker         & 21             & 12   & 6   & 39       \\
Egregor     & 9              & 14   & 3  & 26       \\ 
DarkSide & 3             & 16  & 7    & 26       \\
LockBit 2.0        & 2              & 15  & 6   & 23       \\
SamSam         & 23             & 0  & 1   & 24       \\
Cuba      & 17              & 1    & 1 & 19       \\
Hive      & 1              & 15    & 2 & 18       \\
  \midrule

Total        & 292            & 465    & 256  & 1013      \\
  \bottomrule

\end{tabular} 
}
\end{table}

It is well documented that ransomware families commonly rebrand, most often due to the disbanding of a group~\cite{RansomwareRevenueMore}. Commonly, this analysis looks at the structure of the malware itself or the group's ransom site. Various reports have also identified rebranding based on blockchain analysis, most often in one-off scenarios~\cite{RansomwareRevenueMore,trmlabsAnalysisCorroborates,trmlabsFirstCrypto}. We look to replicate and expand on these results at a much wider scale -- across all payments that we've identified.

To close that gap, we propose a methodology that we call \textit{shared exposure} that we used to identify overlaps between families by tracing shared payment destinations. We compute a weighted multi-hop Jaccard similarity metric between all addresses. That is, given two ransomware payment addresses $a_1$ and $a_2$, we compute $\frac{a_1 \cap a_2}{a_1 \cup a_2}$, where $a_1 \cap a_2$ is the total overlap in funds between the two addresses, and $a_1 \cup a_2$ is the total amount of funds between the two addresses.

\begin{figure}[t]
\centering\includegraphics[width=9cm]{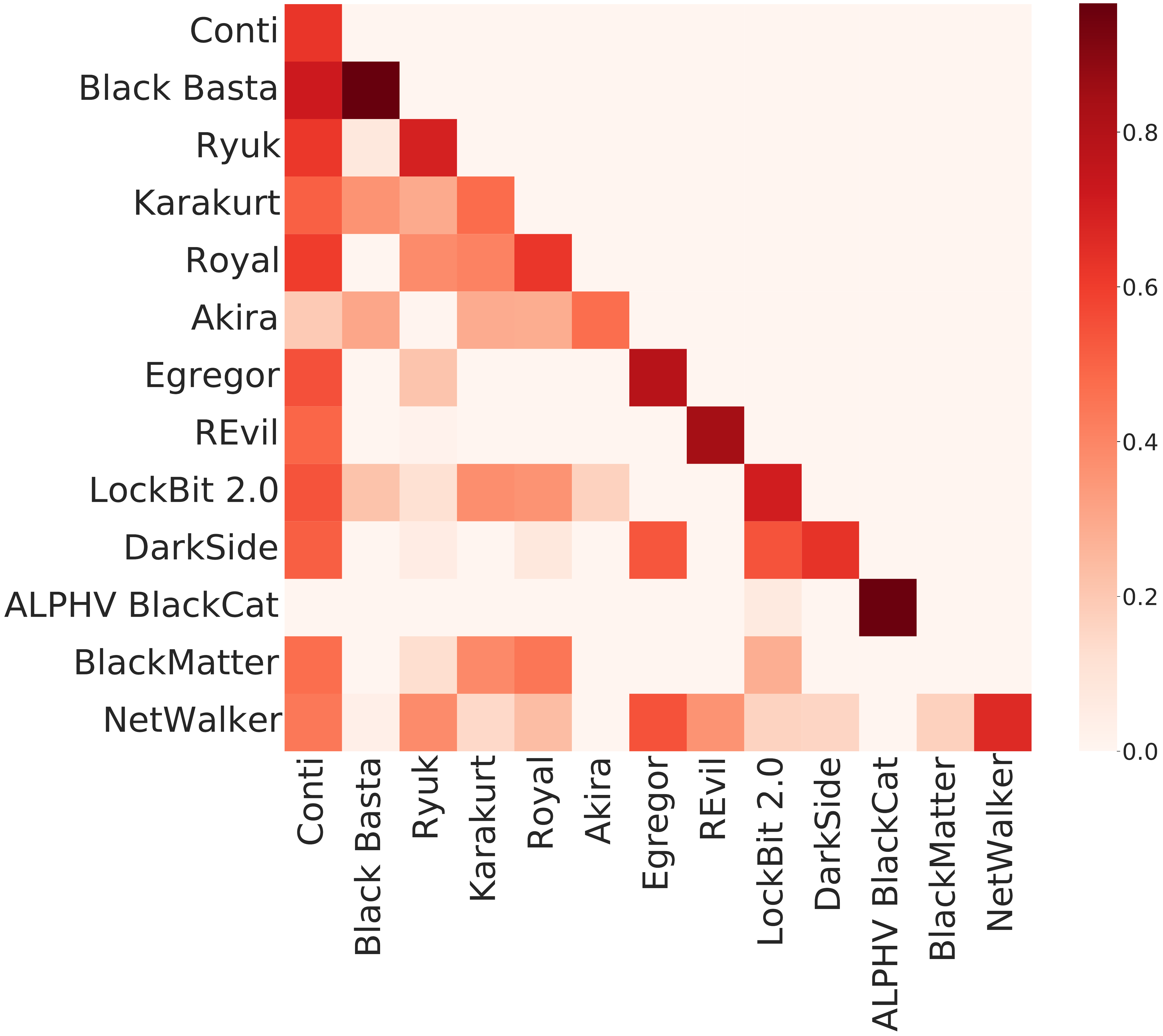}
\caption{Overlap among ransomware families.}
\label{fig:overlap}
\end{figure}

We apply this methodology to both ransomware addresses observed in the wild via the Ransomwhere dataset and to our newly-discovered ransomware addresses. These results are depicted in Figure~\ref{fig:overlap}.

A number of prominent rebrandings between ransomware actors that have been publicly documented are evident. For instance, there is a high degree of overlap among Conti, Black Basta, Ryuk, Karakurt, Royal, and Akira -- illustrated by the cluster of overlaps in the top left corner -- which is consistent with prior reporting of rebrandings among Conti ~\cite{trmlabsFirstCrypto,arctic,trmlabsAnalysisCorroborates}.

Other prominent rebrandings, such as ties between LockBit, DarkSide, ALPHV BlackCat, and BlackMatter are also evident. However, not all ties are as strong -- for instance, ALPHV BlackCat only exhibits slight overlap with LockBit 2.0 (which may be due to a low sample size of addresses for both), and there is little overlap between REvil and LockBit 2.0 despite reports of ties between the two operators. Notably, the majority of ransomware families exhibit high degrees of overlap among themselves. This is not a given, and this helps validate that the labels assigned are indeed accurate.



\subsection{Splitting}

To further understand our dataset, we analyze splitting behavior. Ransomware actors commonly split payments between the core operators and the affiliates, with the affiliate most often receiving the largest portion of the ransom. We leverage the splitting algorithm from Gray et al. in order to identify splitting activity~\cite{Gray2022MoneyOM}. In short, we traverse up to 3 transactions from the payment and look for round split amounts. In total, 745 of the 1,130 addresses exhibited splitting, including 50\% of Ransomwhere addresses, 59\% of original cluster addresses, and 100\% of expanded cluster addresses (due to the heuristics requiring splitting).

\begin{figure}[t]
\includegraphics[width=9cm]{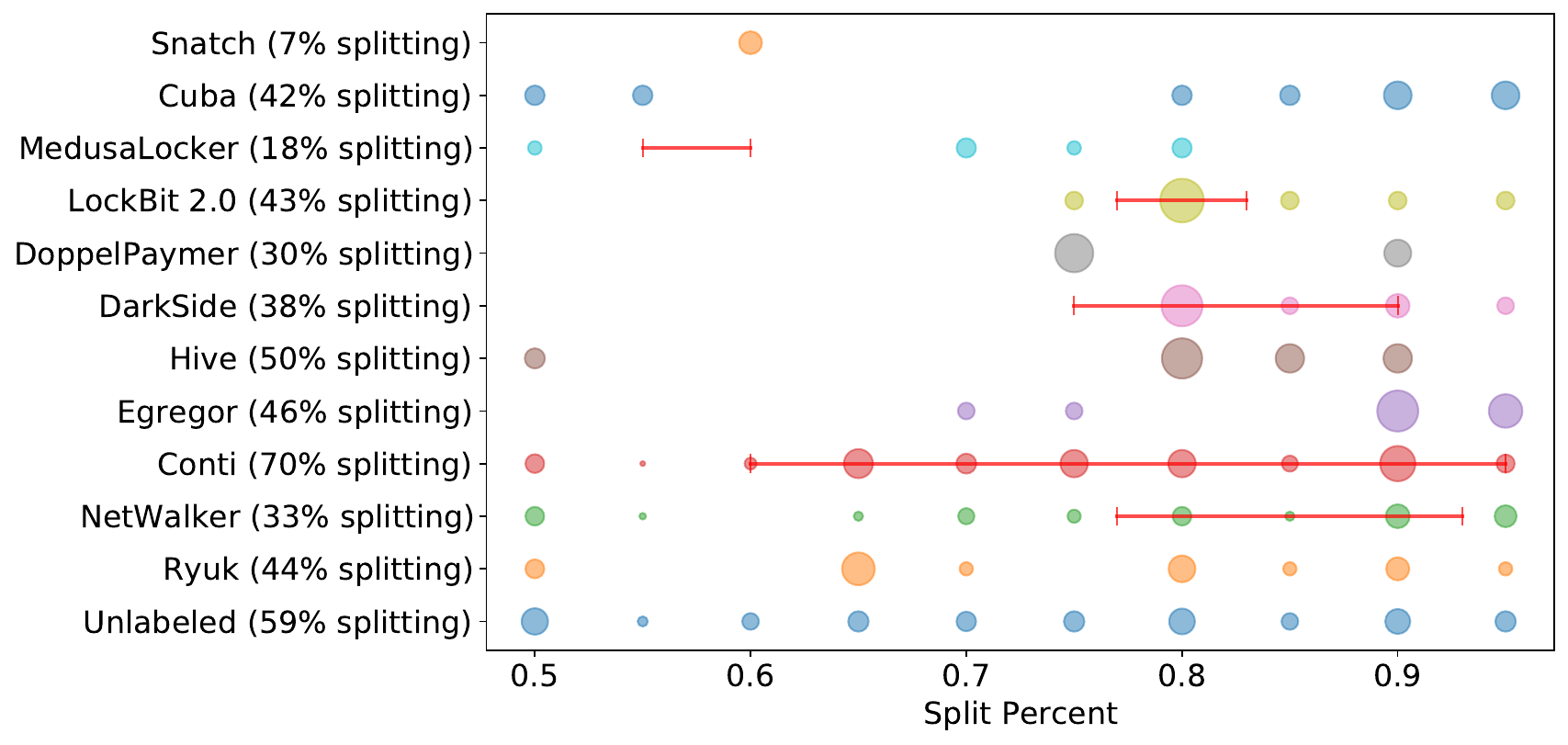}
\caption{Observed splitting behavior by family. Publicly reported splitting percentage ranges are depicted in red~\cite{netskope,mcafee,hhs,Gray2022MoneyOM,googleShiningLight}.}
\label{fig:splits}
\end{figure}

Figure~\ref{fig:splits} shows the splitting behavior across various RaaS families. Certain families, such as Conti and Ryuk, consistently exhibit splitting a majority of the time, while others like Snatch and MedusaLocker only infrequently exhibit splitting. This could either be due to less consistent splitting activity, or by splitting after money has been laundered through a mixer. Additionally, we observe that the split percentage varies by family. Some, such as DarkSide and LockBit, appear to be more generous in allowing their affiliates to keep a greater portion of the ransom payment (typically over 80\%). Others, including Conti and Netwalker, have less consistency in split percents and seem to often give the operators a greater share.

As can be seen in the Figure~\ref{fig:splits}, the split percent often varies widely. Existing public reporting on split percentages for various ransomware groups is scattered. LockBit, for instance, has stated that affiliates receive 80\% of payments while the operator receives 20\%~\cite{netskope}. This appears largely consistent with observed LockBit payments. For other groups, such as Netwalker, sources have reported an 80-90\% split rate~\cite{mcafee}. In reality, Netwalker exhibits much more varied splitting behavior, ranging from 50-95\%.

\begin{figure}[H]
\centering\includegraphics[width=8cm]{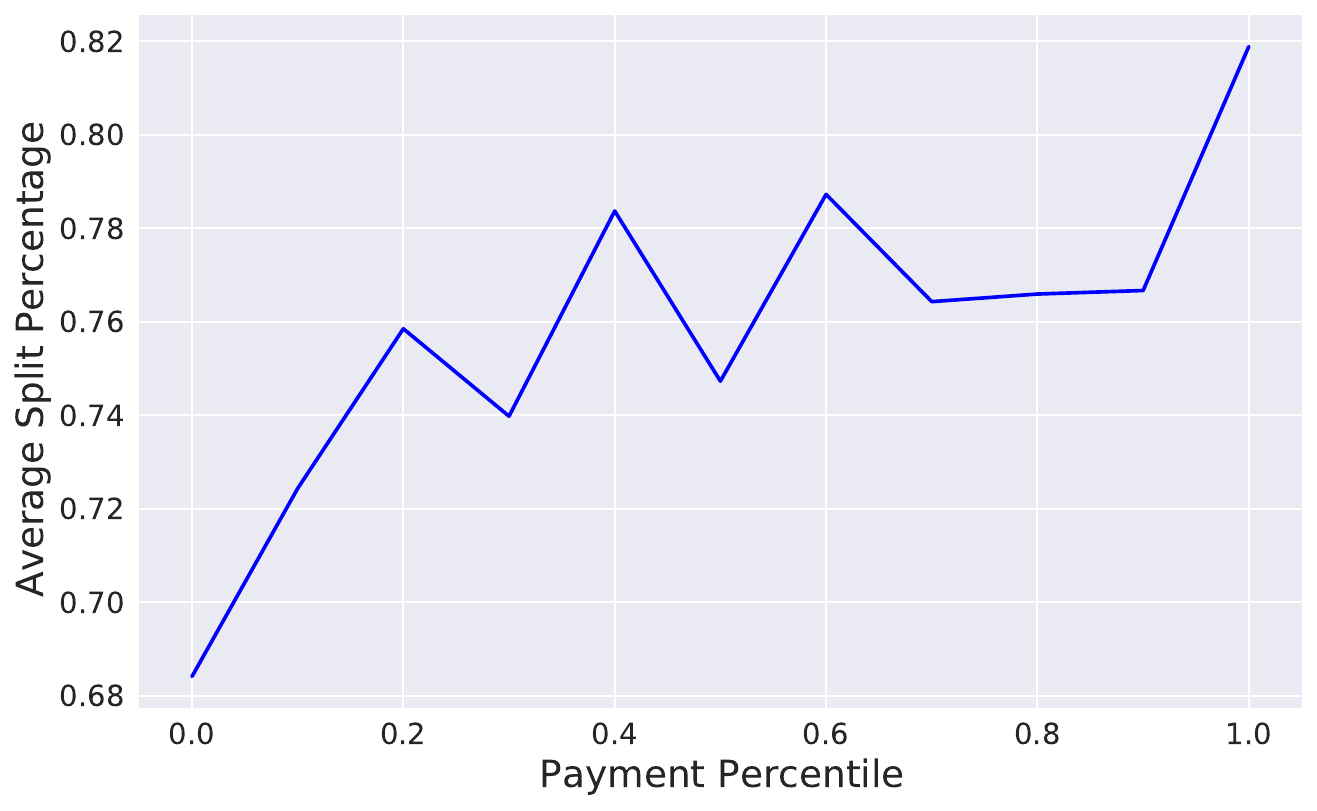}
\caption{Average split percentage by payment percentile among addresses that have exhibited splitting. There appears to be correlation between larger ransomware payments and a higher split percentage kept by affiliates.}
\label{fig:splitPcts}
\end{figure}

We also observe correlation between the payment size and the split percentage. Figure~\ref{fig:splitPcts} shows the average split percentage by payment percentile among addresses that have exhibiting splitting. We note that the average split percentage increases as the payment size increases -- suggesting that ransomware operators might be willing to give affiliates a greater cut of the payment as the amount increases.

\section{Discussion}
\label{discussion}
In our study, we demonstrated the efficacy of applying novel heuristics based on the activities of ransomware negotiators to map ransomware payments at a significantly larger scale than any prior published research. Our research investigates payments and exposure to a ransomware negotiator through the identification of a cluster making ransomware payments, shedding light on the critical role of negotiators and over-the-counter (OTC) desks within the RaaS ecosystem. Tracking adversaries between ransomware groups on the blockchain provides a valuable tool for understanding potential connections between adversaries.

Our heuristics represent a conservative approach to identifying ransomware payments. Additional research can likely further expand on the sets of payments we have identified. For instance, while we have labeled \$283M in payments from Cluster B as ransomware payments, it is entirely possible that the remaining \$50M in payments from Cluster B are also ransomware payments. Likewise, our techniques used to construct our expanded dataset can be repeated given more seed addresses.

One of the primary constraints of applying our heuristics relates to the set of seed addresses used. As both sets of heuristics we develop rely on knowledge of existing ransomware payments, the resulting datasets are inherently limited to the sources of payments used. While we know that ransomware activity has continued in 2023 and 2024, for example, our datasets have limited payments in these years due to a lack of seed payments. We note that Clusters A and B are no longer active. It is possible that there are other, newer clusters used by negotiators that can be identified with more recent payments. Alternatively, negotiators may begin to use different techniques for processing payments to prevent this type of analysis from occurring.

We are sharing the addresses that we identified so that it can support additional research and collaboration within the cybersecurity community. This collective effort can lead to more robust and comprehensive strategies for tracking and mitigating ransomware activities on the blockchain.

In addition to cybersecurity researchers, intermediaries like cryptocurrency exchanges, play a role in the collective effort to mitigate ransomware activities on the blockchain. This research focused on ransomware negotiators and OTC desks, which often interacts with exchanges to source liquidity for large orders~\cite{InstituteSecurityTechnologyMapping}. 

Law enforcement agencies like the FBI discourage victims from paying ransoms, as this fuels the ecosystem and continues to make ransomware a profitable business~\cite{FBIGuidanceEvolves}. However, they also understand the challenge posed by ransomware as critical sectors like healthcare are continually being targeted. If victims decide to pay, they risk sending payments to a sanctioned entity~\cite{MedicalTargetedRansomwareBreaking}. Additionally, there is no guarantee of recovering the encrypted files. Despite these challenges, the FBI will continue to work with victims, offering guidance and support. Cooperation with law enforcement can lead to better outcomes in terms of investigations, mitigations, and reporting. 

More comprehensive reporting of ransomware incidents and associated payments would aid in tracking ransomware operations, and further understanding the ransomware payments ecosystem. In the United States, the Cybersecurity and Infrastructure Security Agency (CISA) is in the process of implementing the Cyber Incident Reporting for Critical Infrastructure Act (CIRCIA). The Act, which was signed into law in March 2022, requires critical infrastructure entities in the United States to report cyber incidents to CISA within 72 hours and ransom payments within 24 hours of payment. The proposed rule includes detailed reporting requirements following an incident such as ransom amount, currency type, payment method, recipient information, and the payment address and transaction identifier. This information may help federal agencies better identify changing ransomware tactics, as well as ransomware payment addresses~\cite{CyberIncidentReporting}.

\section{Related Work}
Cryptocurrency tracing is an essential aspect of understanding and combating ransomware. Various heuristics and methodologies have been developed to trace cryptocurrency transactions, each contributing to the field. This section reviews key techniques and studies relevant to our research methodology, analysis, and conclusions. Our research also included a review of several papers that analyzed the ransomware ecosystem, either through classification, identification, or revenue estimation of ransomware adversaries. These studies are reviewed and included in Section~\ref{subsec:ransomware-payments-analysis}.


Our research leveraged BlockSci, an open-source blockchain analysis platform designed to facilitate the investigation of cryptocurrency transactions~\cite{blocksci}. BlockSci supports efficient querying of blockchain data and employs several key techniques:

\begin{itemize}
  \item \textbf{Multi-Input Clustering:} This heuristic assesses that if multiple inputs in a transaction are used together, they belong to the same entity. Multi-input clustering helps to group addresses that are controlled by the same actor, enabling the tracing of funds through different addresses and transactions~\cite{ron2013quantitative}. 
  \item \textbf{Change Address Detection:}  In Bitcoin transactions, the change address receives leftover funds after the intended payment. The change address is often newly generated and controlled by the sender, helping to determine possible ownership. This heuristic can cluster addresses more accurately, ensuring that all addresses controlled by the same entity are grouped together~\cite{meiklejohnFistfulBitcoinsCharacterizing2013}. 
\end{itemize}

Several studies have applied heuristic analysis to trace illicit cryptocurrency transactions, demonstrating the use of clustering heuristics and address tagging to map out the Bitcoin transaction network. Previous research studies have leveraged the change address heuristic combined with the multi-input heuristics to cluster addresses belonging to the same entity~\cite{meiklejohnFistfulBitcoinsCharacterizing2013}. To further enhance our analysis, in our study we use BlockSci's default multi-input and change address heuristic clustering techniques. The use of BlockSci for clustering allowed us to identify the clusters that most often originate ransomware payments.


Previous research has also focused on mapping ransomware payments. In addition to the research covered in Section~\ref{subsec:ransomware-payments-analysis}, Gomez et al. introduced back-and-forth tracing, which involves traversing the graph of known illicit addresses both forward and backwards to identify other addresses are involved in the same cybercrime campaign~\cite{gomezWatchYourBack2022}. Gomez et al. leverage cluster labels to prevent graph explosion. Our approach shares some similarities with Gomez et al.'s approach, although our development of heuristics unique to ransomware negotiators allows us to classify transactions that would be filtered out (due to being labeled as an exchange).




\section{Conclusion}
\label{conclusion}
We have presented a method for identifying ransomware payments and proposed several metrics to evaluate our results. Leveraging prior public datasets of ransomware payments, we used our methods to expand these and we identified over \$700 million in likely ransomware payments. Our data has been made public and, to the best of our knowledge, is the largest public dataset of likely RaaS payments. 

We performed reproducible analyses on our public dataset that largely confirm the scale of magnitude of RaaS payments from proprietary reports. We believe that reproducible methodologies and comprehensive open datasets of ransomware payments are essential for continuing to understand ransomware and evaluating interventions.

\section*{Acknowledgment}

We thank the anonymous reviewers for their insightful and constructive suggestions and feedback, and Crystal for providing access to their platform. Funding for this
work was provided in part by National Science Foundation grants 1844753 and 2039693.

\printbibliography

@inproceedings{meiklejohnFistfulBitcoinsCharacterizing2013,
  title = {A Fistful of Bitcoins: Characterizing Payments among Men with No Names},
  shorttitle = {A Fistful of Bitcoins},
  booktitle = {Proceedings of the 2013 Conference on {{Internet}} Measurement Conference},
  author = {Meiklejohn, Sarah and Pomarole, Marjori and Jordan, Grant and Levchenko, Kirill and McCoy, Damon and Voelker, Geoffrey M. and Savage, Stefan},
  year = {2013},
  month = oct,
  pages = {127--140},
  publisher = {{ACM}},
  address = {{Barcelona Spain}},
  doi = {10.1145/2504730.2504747},
  urldate = {2022-09-17},
  isbn = {978-1-4503-1953-9},
  langid = {english},
}

@dataset{ransomwhere,
  author       = {Cable, Jack},
  title        = {{Ransomwhere: A Crowdsourced Ransomware Payment 
                   Dataset}},
  month        = may,
  year         = 2022,
  publisher    = {Zenodo},
  version      = {1.0.1},
  doi          = {10.5281/zenodo.6562484},
  url          = {https://doi.org/10.5281/zenodo.6562484}
}

@inproceedings{blocksci,
author = {Kalodner, Harry and M\"{o}ser, Malte and Lee, Kevin and Goldfeder, Steven and Plattner, Martin and Chator, Alishah and Narayanan, Arvind},
title = {BlockSci: design and applications of a blockchain analysis platform},
year = {2020},
isbn = {978-1-939133-17-5},
publisher = {USENIX Association},
address = {USA},
booktitle = {Proceedings of the 29th USENIX Conference on Security Symposium},
articleno = {153},
numpages = {18},
series = {SEC'20}
}

@ARTICLE {10308425,
author = {Z. Baig and S. Mekala and S. Zeadally},
journal = {IT Professional},
title = {Ransomware Attacks of the COVID-19 Pandemic: Novel Strains, Victims, and Threat Actors},
year = {2023},
volume = {25},
number = {05},
issn = {1941-045X},
pages = {37-44},
doi = {10.1109/MITP.2023.3297085},
publisher = {IEEE Computer Society},
address = {Los Alamitos, CA, USA},
month = {sep}
}

@misc{mcafee,
	author = {{ATR Operational Intelligence Team}},
	title = {{T}ake a “{N}et{W}alk” on the {W}ild {S}ide},
	howpublished = {\url{https://www.mcafee.com/blogs/other-blogs/mcafee-labs/take-a-netwalk-on-the-wild-side/}},
	year = {},
	note = {[Accessed 03-07-2024]},
}

@article{Gray2022MoneyOM,
  title={{Money Over Morals: A Business Analysis of Conti Ransomware}},
  author={Ian W. Gray and Jack Cable and Benjamin Brown and Vlad Cuiujuclu and Damon McCoy},
  journal={2022 APWG Symposium on Electronic Crime Research (eCrime)},
  year={2022},
  pages={1-12},
  url={https://api.semanticscholar.org/CorpusID:258298293}
}

@misc{gomezWatchYourBack2022,
  title = {Watch {{Your Back}}: {{Identifying Cybercrime Financial Relationships}} in {{Bitcoin}} through {{Back-and-Forth Exploration}}},
  shorttitle = {Watch {{Your Back}}},
  author = {Gomez, Gibran and {Moreno-Sanchez}, Pedro and Caballero, Juan},
  year = {2022},
  month = oct,
  number = {arXiv:2206.00375},
  eprint = {2206.00375},
  primaryclass = {cs},
  publisher = {{arXiv}},
  urldate = {2023-01-08},
  archiveprefix = {arxiv},

}

@misc{congAnatomyCryptoEnabledCybercrimes2022a,
	address = {Rochester, NY},
	type = {{SSRN} {Scholarly} {Paper}},
	title = {An {Anatomy} of {Crypto}-{Enabled} {Cybercrimes}},
	url = {https://papers.ssrn.com/abstract=4188661},
	doi = {10.2139/ssrn.4188661},
	language = {en},
	urldate = {2022-08-24},
	author = {Cong, Lin William and Harvey, Campbell R. and Rabetti, Daniel and Wu, Zong-Yu},
	month = jul,
	year = {2022},
}

@article{contiEconomicSignificanceRansomware2018a,
	title = {On the {Economic} {Significance} of {Ransomware} {Campaigns}: {A} {Bitcoin} {Transactions} {Perspective}},
	volume = {79},
	issn = {01674048},
	shorttitle = {On the {Economic} {Significance} of {Ransomware} {Campaigns}},
	url = {http://arxiv.org/abs/1804.01341},
	doi = {10.1016/j.cose.2018.08.008},
	language = {en},
	urldate = {2022-07-18},
	journal = {Computers \& Security},
	author = {Conti, Mauro and Gangwal, Ankit and Ruj, Sushmita},
	month = nov,
	year = {2018},
	note = {arXiv:1804.01341 [cs]},
	pages = {162--189},
}

@article{paquet-cloustonRansomwarePaymentsBitcoin2019a,
	title = {Ransomware payments in the {Bitcoin} ecosystem},
	volume = {5},
	issn = {2057-2085, 2057-2093},
	url = {https://academic.oup.com/cybersecurity/article/doi/10.1093/cybsec/tyz003/5488907},
	doi = {10.1093/cybsec/tyz003},
	language = {en},
	number = {1},
	urldate = {2022-07-12},
	journal = {Journal of Cybersecurity},
	author = {Paquet-Clouston, Masarah and Haslhofer, Bernhard and Dupont, Benoît},
	month = jan,
	year = {2019},
	note = {Number: 1},
	pages = {tyz003},
}

@inproceedings{gomezCybercrimeBitcoinRevenue2023b,
	title = {Cybercrime {Bitcoin} {Revenue} {Estimations}: {Quantifying} the {Impact} of {Methodology} and {Coverage}},
	shorttitle = {Cybercrime {Bitcoin} {Revenue} {Estimations}},
	url = {http://arxiv.org/abs/2309.03592},
	doi = {10.1145/3576915.3623094},
	language = {en},
	urldate = {2024-02-11},
	booktitle = {Proceedings of the 2023 {ACM} {SIGSAC} {Conference} on {Computer} and {Communications} {Security}},
	author = {Gomez, Gibran and van Liebergen, Kevin and Caballero, Juan},
	month = nov,
	year = {2023},
	note = {arXiv:2309.03592 [cs]},
	pages = {3183--3197},
}

@article{gautschiFindingAnalyzingLockbit,
	title = {Finding and {Analyzing} {Lockbit} {Split} {Transactions} on the {Bitcoin} {Blockchain}},
	language = {en},
	author = {Gautschi, Alain},
}

@misc{googleShiningLight,
	author = {Jordan Nuce and Jeremy Kennelly and Kimberly Goody and Andrew Moore and Alyssa Rahman and Matt Williams and Brendan McKeague and Jared Wilson},
	title = {{S}hining a {L}ight on {D}{A}{R}{K}{S}{I}{D}{E} {R}ansomware {O}perations | {G}oogle {C}loud {B}log --- cloud.google.com},
	howpublished = {\url{https://cloud.google.com/blog/topics/threat-intelligence/shining-a-light-on-darkside-ransomware-operations/}},
	year = {}
}

@misc{hhs,
	author = {{Department of Health and Human Services}},
	title = {MedusaLocker Ransomware},
	howpublished = {\url{https://www.hhs.gov/sites/default/files/medusalocker-ransomware-analyst-note.pdf}},
	year = {}
}

@misc{netskope,
	author = {Gustavo Palazolo},
	title = {{N}etskope {T}hreat {C}overage: {L}ock{B}it’s {R}ansomware {B}uilder {L}eaked --- netskope.com},
	howpublished = {\url{https://www.netskope.com/blog/netskope-threat-coverage-lockbits-ransomware-builder-leaked}},
	year = {}
}

@misc{trmlabsAnalysisCorroborates,
	author = {{TRM Labs}},
	title = {{T}{R}{M} {A}nalysis {C}orroborates {S}uspected {T}ies {B}etween {C}onti and {R}yuk {R}ansomware {G}roups and {W}izard {S}pider},
	howpublished = {\url{https://www.trmlabs.com/post/analysis-corroborates-suspected-ties-between-conti-and-ryuk-ransomware-groups-and-wizard-spider}},
	year = {}
}

@misc{arctic,
	author = {Steven Campbell and Akshay Suthar and Connor Belfiore},
	title = {{C}onti and {A}kira: {C}hained {T}ogether},
	howpublished = {\url{https://arcticwolf.com/resources/blog/conti-and-akira-chained-together/}},
	year = {}
}

@misc{trmlabsFirstCrypto,
	author = {{TRM Labs}},
	title = {{T}he {F}irst {C}rypto {W}ar? {A}ssessing the {I}llicit {B}lockchain {E}cosystem {O}ne {Y}ear {I}nto {R}ussia's {I}nvasion of {U}kraine},
	howpublished = {\url{https://www.trmlabs.com/post/the-first-crypto-war-assessing-the-illicit-blockchain-ecosystem-one-year-into-russia-ukraine-war}},
	year = {}
}

@misc{coveware,
	title = {{RaaS devs hurt their credibility by cheating affiliates in Q1 2024}},
	url = {https://www.coveware.com/blog/2024/4/17/raas-devs-hurt-their-credibility-by-cheating-affiliates-in-q1-2024},
	urldate = {2024-04-17},
}

@misc{RansomwareRevenueMore,
	title = {Ransomware {Revenue} {Down} {As} {More} {Victims} {Refuse} to {Pay} - {Chainalysis}},
	url = {https://www.chainalysis.com/blog/crypto-ransomware-revenue-down-as-victims-refuse-to-pay/},
	urldate = {2024-02-25},
}

@misc{RansomwareHitBillion,
	title = {Ransomware {Hit} \$1 {Billion} in 2023},
	url = {https://www.chainalysis.com/blog/ransomware-2024/},
	urldate = {2024-02-25},
}

@misc{genesis,
	title = {{Genesis to Shutter Crypto Trading Desk for U.S. Market}},
	url = {https://www.coindesk.com/business/2023/09/05/genesis-global-trading-to-shutter-crypto-spot-trading-desk/},
	urldate = {2023-09-05},
author={Danny Nelson}
}

@misc{oosthoek2022tale,
      title={A Tale of Two Markets: Investigating the Ransomware Payments Economy}, 
      author={Kris Oosthoek and Jack Cable and Georgios Smaragdakis},
      year={2022},
      eprint={2205.05028},
      archivePrefix={arXiv},
      primaryClass={cs.CR}
}

@misc{ransomware_mass_market, 
    title={Ransomware goes mass market}, 
    url={https://www.chainalysis.com/blog/ransomware-raas-cryptocurrency-2019/}, 
    journal={Chainalysis}, 
    author={Team, Chainalysis}, 
    year={2020}, 
    month={Jan}
}

@article{RansomwareNegotiations2021,
    author={Rachel Monroe},
    title={How to Negotiate with Ransomware Hackers},
    journal={The New Yorker},
    year={2021},
    month={June},
    url={https://www.newyorker.com/magazine/2021/06/07/how-to-negotiate-with-ransomware-hackers},
    note={Accessed: 2024-04-19}
}

@misc{MediumOTC2020,
    title={How Does Crypto OTC Actually Work?},
    author={Connor Dempsey},
    howpublished={Medium},
    year={2020},
    url={https://medium.com/circle-research/how-does-crypto-otc-actually-work-e2215c4bb13},
    note={Accessed: 2024-04-19}
}

@misc{Chainalysis2021Crypto,
	title = {The {Chainalysis} 2021 {Crypto} {Crime} {Report}},
	url = {https://go.chainalysis.com/2021-Crypto-Crime-Report-demo.html},
	urldate = {2024-06-21},
}

@misc{cybersecuritydiveHiveTakedown,
	author = {Matt Kapko},
	title = {{H}ive takedown puts ‘small dent’ in ransomware problem --- cybersecuritydive.com},
	howpublished = {\url{https://www.cybersecuritydive.com/news/hive-takedown-small-dent-ransomware/642066/}},
	year = {}
}

@misc{chainalysisOFACSanctions,
	author = {{Chainalysis Team}},
	title = {{O}{F}{A}{C} {S}anctions {T}racker: {H}ow {S}anctions {I}mpact {C}rypto {C}rime - {C}hainalysis --- chainalysis.com},
	howpublished = {\url{https://www.chainalysis.com/blog/ofac-sanctions/}},
	year = {}
}

@misc{RansomwareGoesMassa,
	title = {Ransomware {Goes} {Mass} {Market} - {Chainalysis}},
	url = {https://www.chainalysis.com/blog/ransomware-raas-cryptocurrency-2019/},
	urldate = {2024-06-23},
}

@misc{RansomwareSkyrocketed2020,
	title = {Ransomware {Skyrocketed} in 2020, {But} {There} {May} {Be} {Fewer} {Culprits} {Than} {You} {Think} - {Chainalysis}},
	url = {https://www.chainalysis.com/blog/ransomware-ecosystem-crypto-crime-2021/},
	urldate = {2024-06-23},
}

@misc{RansomwareRevenueMorea,
	title = {Ransomware {Revenue} {Down} {As} {More} {Victims} {Refuse} to {Pay} - {Chainalysis}},
	url = {https://www.chainalysis.com/blog/crypto-ransomware-revenue-down-as-victims-refuse-to-pay/},
	urldate = {2024-06-23},
}

@misc{RansomwareHitBillionb,
	title = {Ransomware {Hit} \$1 {Billion} in 2023},
	url = {https://www.chainalysis.com/blog/ransomware-2024/},
	urldate = {2024-06-23},
}

@inproceedings{ron2013quantitative,
  title={Quantitative analysis of the full bitcoin transaction graph},
  author={Ron, Dorit and Shamir, Adi},
  booktitle={International Conference on Financial Cryptography and Data Security},
  pages={6--24},
  year={2013},
  organization={Springer}
}

@misc{FBIGuidanceEvolves,
	title = {{FBI} {Guidance} {Evolves} on {Ransomware} {Payments} {\textbar} {Decipher}},
	url = {https://duo.com/decipher/fbi-guidance-evolves-on-ransomware-payments},
	urldate = {2024-07-06},
}

@misc{CyberIncidentReporting,
	title = {Cyber {Incident} {Reporting} for {Critical} {Infrastructure} {Act} of 2022 ({CIRCIA}) {\textbar} {CISA}},
	url = {https://www.cisa.gov/topics/cyber-threats-and-advisories/information-sharing/cyber-incident-reporting-critical-infrastructure-act-2022-circia},
	urldate = {2024-07-06},
}

@misc{MedicalTargetedRansomwareBreaking,
	title = {Medical-{Targeted} {Ransomware} {Is} {Breaking} {Records} {After} {Change} {Healthcare}’s \${22M} {Payout} {\textbar} {WIRED}},
	url = {https://www.wired.com/story/change-healthcare-22-million-payment-ransomware-spike/},
	urldate = {2024-07-06},
}

@misc{InstituteSecurityTechnologyMapping,
	title = {Institute for {Security} and {TechnologyMapping} the {Ransomware} {Payment} {Ecosystem}: {A} {Comprehensive} {Visualization} of the {Process} and {Participants} - {Institute} for {Security} and {Technology}},
	url = {https://securityandtechnology.org/virtual-library/reports/mapping-the-ransomware-payment-ecosystem-a-comprehensive-visualization-of-the-process-and-participants/},
	urldate = {2024-07-06},
}


\end{document}